
\
\input phyzzx
\catcode`\@=11

\def\eqaligntwo#1{\null\,\vcenter{\openup\jot\m@th
\ialign{\strut\hfil
$\displaystyle{##}$&$\displaystyle{{}##}$&$\displaystyle{{}##}$\hfil
\crcr#1\crcr}}\,}
\catcode`\@=12
\overfullrule = 0 pt
\normalbaselineskip  = 20pt plus 0.2pt minus 0.1pt
\hsize = 6.0 in
\hoffset =  0.25 in

\def\tr{{\rm Tr\ }}

\def\z{{\cal Z}}

\def\la{\lambda_a}
\def\lb{\lambda_b}

\def\mua{\mu_a}
\def\mub{\mu_b}

\def\ref{\REF}
\def\cz{{\cal Z}}
\def\cd{{\cal D}}
\def\pa{\partial}
\def\N{{1\over N}}
\def\sig{\sigma^{}}

\def\pak{\pa_k}
\def\paj{\pa_j}
\def\gg{{\textstyle {1\over3}g}}
\def\ee{\eqn\eq}
\def\ts{\textstyle}
\def\fa{F_a}
\def\fb{F_b}

\def\NP{{\it Nucl. Phys.\ }}

\def\PL{{\it Phys. Lett.\ }}

\def\PRL{{\it Phys. Rev. Lett.\ }}
\def\CMP{{\it Comm. Math. Phys.\ }}

\def\IJMP{{\it Int. Jour. Mod. Phys.\ }}
\def\Mod{{\it Mod. Phys. Lett.\ }}
\def\MPL{\Mod}

\Pubnum={PUPT-1282}
\date = { December 4, 1991}
\titlepage
\overfullrule = 0 pt
\normalbaselineskip  = 20pt plus 0.2pt minus 0.1pt
\hsize = 6.0 in
\hoffset =  0.25 in
\title
{
 Unitary And Hermitian  Matrices In An External Field II: The Kontsevich
Model And Continuum Virasoro Constraints.}
\author{David J. Gross\foot{Research supported in part by NSF grant
PHY90-21984}
 and Michael J. Newman.  }
\address
    {
    Joseph Henry Laboratories, Department of Physics,
    Princeton University, Princeton, N.J. 08544\foot{\rm E-mail:
gross@puhep1.princeton.edu; newman@puhep1.princeton.edu}
    }

\abstract{
We give a simple derivation of the Virasoro constraints in the
Kontsevich model, first derived by Witten.
We generalize  the method to a  model of unitary matrices,
for which we
 find a new set of Virasoro constraints. Finally we discuss the
solution for symmetric
matrices in an external field.}

\endpage

\chapter{Introduction.}

There has recently been considerable progress in the study of matrix
models,
 following the
\ref\kont{M. Kontsevich, ``Intersection theory on the moduli space of
curves,'' preprint (1990).}
remarkable proof by Kontsevich that the intersection numbers (correlation
functions) of two-dimensional
topological gravity are generated by a  new type of
\ref\witten{E. Witten, ``On the Kontsevich model and other models of two
dimensional gravity,'' IAS preprint IASSNS-HEP-91/24.}
matrix model~[\kont]. A short time later, Witten showed~[\witten]
that the partition function of this model obeys the Virasoro
\ref\dvv{R. Dijkgraaf, E. Verlinde, and H. Verlinde, \NP {\bf B348}
(1991) 435.}
\ref\fkn{M. Fukuma, H. Kawai, and R. Nakayama, \IJMP {\bf A6} (1991)
1385.}
constraints of the  one-matrix model~[\dvv,~\fkn],
thus completing the chain of arguments in
\ref\wittop{E. Witten, \NP {\bf B340} (1990) 281.}
a  proof of the old conjecture that topological
gravity and matrix models are equivalent~[\wittop].

However, Witten's  lengthy
proof depends on a cumbersome diagrammatic expansion.
We present in section~2 a
much simpler derivation, which arose out of recent work into matrix
\ref\us{D. J. Gross and M. J. Newman,
\PL {\bf 266B} (1991) 291.}
integrals involving an external field~[\us].\foot{See also
\ref\MakSem{Yu.~Makeenko and G. W. Semenoff, ``Properties of Hermitean
matrix models in an external field,'' ITEP/University of British
Columbia preprint (July 1991).}
 ref.~\MakSem.} This approach  allows us
to consider other models;
 and in
   section~3  we tackle  the case  of
\ref\BG{E.~Br\'ezin and D.~J.~Gross, \PL {\bf 97B} (1980) 120.}
\ref\PSA{V.~Periwal and D.~Shevitz, \PRL {\bf 64} (1990) 1326.}
\ref\PSB{V.~Periwal and D.~Shevitz, \Mod {\bf A5} (1990) 1147.}
 a unitary matrix in an external field~[\BG, \us],
 deriving a set of Virasoro
constraints which  describe the continuum limit of the multicritical
unitary matrix models~[\PSA,~\PSB]. These results are  new, but after
this work was completed, we learned\foot{We
thank E. Witten for informing us of this.}  that
\ref\holl{T. Hollowood, L. Miramontes, A. Pasquinucci, and C. Nappi,
Princeton/IAS preprint IASSNS-HEP-91/59, PUPT-1280.}
another group~[\holl] has found the same constraints from the
\ref\CDMB{\v C. Crnkovi\'c, M. Douglas, and G. Moore, ``Loop equations
and the topological phase of multicritical matrix models,'' Yale/Rutgers
University preprint YCTP-P25-91/RU-91-36 (August 1991).}
complementary viewpoint of the mKdV flows~[\PSB,~\CDMB].

In section~4 we present the results of some work on symmetric matrices
in an external field. As far as we know this model has never before been
investigated, so we stray somewhat from the main subject of this paper
and give some general details. Unfortunately this solution
is incomplete, and in particular we fail to produce for symmetric
matrices any results
generalizing the Virasoro constraints of sections~2 and~3.

In the final section
 we discuss our results, and speculate what these
might teach us about unitary matrix integrals.

\chapter{Virasoro constraints from Schwinger-Dyson equations.}

Consider the  integral over $N\times N$ hermitian matrices
$$\hat\cz =
 \int \cd \hat M \exp \tr\!\!\left(X\hat M
-\gg \hat M^3\right) \ , \eqn\dzhat
$$
which is a function of the $N$ eigenvalues $x_a$ of the hermitian matrix
$X$.\foot{Note that for the purposes of this paper, we define the integral
without an $N$ in the exponent.}
\ref\Kostov{I.~Kostov, in {Jaca 1988, Proceedings,} {\it Non-perturbative
aspects of the standard model},  295.}
In refs.~\us,~\Kostov, it is shown how to evaluate $\hat\cz$ using
the Schwinger-Dyson equations of motion,
$$\int \cd \hat M {\pa\over\pa\hat M}\exp \tr\!\!\left(X\hat M
-\gg \hat M^3\right) = \left( X^{T}
 -g{\pa^2\over\pa X^2}\right)  \hat\cz =0 \ , \eqn\eq $$
which can be recast as a set of $N$ differential equations in terms of the
eigenvalues $x_a$,
$$
 {\pa^2 \hat \cz\over\pa x_a^2} + \sum_{b\neq a} {1\over
x_a-x_b}\left({\pa \hat \cz\over \pa x_a}-{\pa \hat \cz\over \pa x_b}
\right)  ={1\over g} x_a \hat \cz \ . \eqn\zhat
$$
Full details of the derivation are given in refs.~\BG,~\us.
In ref.~\Kostov\ these equations were used to find $\hat\cz$ in the
spherical approximation, and recently we showed how to
 extend this solution to
all orders in $N$~[\us].\foot{The reader is advised that an early
preprint version of ref.~\us\ contained some mistakes in the
introduction.}
 In this paper we start from these same equations, but the analysis will
be very different.

Closely related to $\hat\cz$ is the  integral
$$\cz =\int\cd M \exp \tr\!\!\left(-\half AM^2-\gg M^3\right) \ , \eqn\kint
$$
for a positive definite  hermitian matrix $A$.
In his ground-breaking paper~[\kont], Kontsevich showed how the expansion of
$\cz$ in terms of Feynman diagrams can be interpreted as
 a cell decomposition of the
\ref\penner{R. Penner, \CMP (1987); {\it J.~Diff.~Geom.} {\bf 27} (1988) 35.}
\ref\harer{J. Harer, {\it Inv.~Math.} {\bf 84} (1986) 157.}
\ref\bowditch{B. H. Bowditch and D. B. A. Epstein, {\it Topology} {\bf
27} (1988) 35.}
moduli space of Riemann surfaces [\penner,~\harer,~\bowditch].
By a shift of  integration variables, it is easy to see that
 $\hat\cz$ and  $\cz$ are related by
$$\hat\cz =
   \exp\tr{A^3\over 12g^2}  \times \cz
\ , \eqn\zrel
$$
provided we choose $X={A^2\over  4g}$.
Our goal now is to use~\zhat\ and~\zrel\ to show that
$\cz$ obeys the Virasoro constraints of the one-matrix model~[\dvv,~\fkn].

Following Kontsevich, we factor $\cz$ in the form
$$\cz =
\prod_{a,b}\left(\mua+\mub\right)^{-1/2} Y \ , \eqn\zfact
$$
where $\{\mua\}$ are the eigenvalues of $A$.
Then, after a change of variables to $\la\equiv\mua^2=4gx_a$,
eqns.~\zhat\ and~\zrel\
imply that
$Y$ satisfies the set of differential equations
$${\pa^2 Y\over \pa\la^2} +{\pa Y\over\pa\la} \left({\mua\over4g^2}-
{Z_a\over\mua}\right) + \sum_{b\neq a}
{1\over\mub^2-\mua^2}
\left({\pa Y\over\pa\lb} - {\pa Y\over\pa\la} \right)
 +Y\left({1\over16\mua^4}+{t_0^2\over4\mua^2}\right) = 0 \ ,
\eqn\eq
$$
where we have defined
$$Z_a=\sum_b{1\over\mua+\mub}; \qquad t_k={-1\over
2k+1}\sum_b{1\over\mub^{2k+1}} \ , \quad k\geq 0 \ . \eqn\eq
$$
Next we make good use of  Kontsevich's important  observation
  that  the  perturbative expansion of $Y$ depends on the eigenvalues of
$A$ only through the invariants
$t_k$.
This lets us
change variables  from $\la$ to
$t_k$, \ie,
$${\pa\over\pa\la}=\half\sum_k {1\over\mua^{2k+3}} {\pa\over\pa t_k} \ ,
\eqn\eq
$$
  and after some algebra  we  obtain
$$\eqalign{&{\ts{ 1\over4}}\sum_{k,j=0}^\infty\ {1\over\mua^{2k+2j+6}}\
\pak\paj Y
+{1\over8g^2}\sum_{k=0}^\infty {1\over\mua^{2k+2}}\pak Y \cr
&+\half\sum_{k=0}^\infty \pak Y\left({(2k+3)t_{k+1}\over\mua^2} +
{(2k+1)t_k\over\mua^4} + \cdots {t_0\over\mua^{2k+4}} \right)
+{1\over16\mua^4} Y + {t_0^2\over 4\mua^2} Y =0 \ , }\eqn\eq
$$
where $\pak\equiv \pa/\pa t_k$.
Note that these equations are formally valid independent of the large-$N$
limit, except that only for $N\to \infty$ are the $t_k$'s truly
independent, and for finite $N$ the sums over $k$ must be
truncated.\foot{For related comments, see section 1.1 of
ref.~\witten.}

To make contact with Kontsevich's results, we
 set $g=i/2$. Then
 we can  pick off the coefficient of $1/\mua^{4+2n}$ to obtain the
equations
$$\eqalign{ n=-1 :&\qquad \sum_{k=1}^\infty (k+\half)t_k\pa_{k-1} Y + {\ts
{1\over 4}} t_0^2 Y = \half \pa_0 Y \ ,\cr
n=0~~ :&\qquad \sum_{k=0}^\infty (k+\half) t_k\pak Y +{\ts
{1\over 16}}Y =\half\pa_1Y
\ , \cr
n\geq1~~ :&\qquad \sum_{k=0}^\infty (k+\half) t_k\pa_{k+n}Y +
{\ts{ 1\over4}}\sum_{k=1}^{n} \pa_{k-1}\pa_{n-k} Y = \half \pa_{n+1} Y \ . }
\eqn\eq
$$
Given that we have used a different normalization for the $t_k$'s,
 these are precisely the equations first derived for the Kontsevich model
in ref.~\witten,
$$L_n Y =\half \pa_{n+1} Y, \qquad n\geq -1 \ , \eqn\eq
$$
\ie, the Virasoro constraints of refs.~\dvv,~\fkn, corresponding to
perturbations
about  the $m=1$ ``topological''  point of the
one-matrix model.
It is remarkable that the entire set of multicritical potentials
should be accessible from the simple cubic potential of~\kint.
\ref\Kaz{V.~A.~Kazakov, \NP {\bf B354} (1991) 614.}
(On the other hand, it is becoming increasingly apparent
that matrix models involving an external matrix field possess a rich
multicritical structure~[\us,~\Kaz].)

Finally, we complete the correspondence between the Kontsevich model and
 topological gravity by
noting that the familiar selection rule for correlation functions  of
scaling operators at genus $g$
$$
\langle \prod_k \tau_k^{n_k}\rangle_g \equiv \prod_k \left({\pa\over\pa
t_k}\right)^{n_k} \left[ \log Y \right]_g   \ee
$$
 (where $\left[ \cdots \right]_g $ means  the contribution at genus
$g$), namely
$$\sum_k n_k(k-1) =3g-3 \ , \eqn\hselr\
$$ is
readily extracted from results in the appendix of ref.~\us. As is well
known,
eqn.~\hselr\ and  the $L_0$ constraint are all one needs to find the
string
susceptibility and
 the scaling dimensions of operators.

\chapter{Unitary matrices.}

The success of the analysis so far suggests that one should investigate
other models involving an external field. One such case,
considered in another context in
refs.~\BG,~\us, is that of a unitary matrix,
$$\bar\cz=\int\cd U \exp\tr\!\!\left( A^\dagger U + U^\dagger A\right) \ ,
\eqn\umat
$$
where $A$ is  now an arbitrary matrix.
As mentioned earlier, we know of  no simple geometrical interpretation for
such an integral.
One approach is to write the unitary matrix in terms of a hermitian
\ref\Neu{H. Neuberger, \NP {\bf B340} (1990) 703.}
\ref\bowick{M.~Bowick, A. Morozov, and D. Shevitz, \NP {\bf B354} (1991)
496.}
matrix, for example  as $U=e^{iH}$~[\Neu], or $U=(1+iH)/(1-iH)$~[\bowick], but
neither of these has a particularly attractive expansion in terms of
surfaces.

The partition function $\bar\cz$ satisfies the differential equations
$${ \partial^2 \bar\z \over \partial A^{~}_{ab} \partial A^\dagger_{bc} }
= \delta_{ac} \bar \z \ . \eqn\diffe $$
After changing variables in these equations to the eigenvalues
$\la\equiv \mua^2$ of $A^\dagger A$, we get the
 Schwinger-Dyson equations for
$\bar\cz$,
$$
{\pa^2  \bar\cz\over\pa \la^2} + \sum_{b\neq a} {1\over
\la-\lb}\left({\pa \bar \cz\over \pa \la}-{\pa \bar \cz\over \pa \lb}
\right) ={1\over\la}\left(1-\sum_b{\pa\bar \cz\over \pa \lb}\right) \ .
\eqn\zbar
$$
Guided  by experience, we expect that we should factor
$\bar\cz$ along the lines of~\zfact; and indeed,
after a small amount of trial and error, we find it  useful to
write
$$ \bar\cz =\prod_{a,b}\left(\mua+\mub\right)^{-1/2}\exp
\Big(2\sum_b\mub\Big) \ \bar Y \ . \eqn\factor
$$
This factor
is what one gets  by   expanding $U$
about the saddle point of the action  as the exponential of a hermitian matrix
and dropping terms higher than
quadratic, and in a sense~\factor\ is the natural analog of the
factorization~\zfact.

It was seen in ref.~\us\ that $\bar Y$ shares with $Y$ the
property of  depending  only on the $t_k$'s. This encourages us to try
changing variables in~\zbar, and we find that
 $\bar Y$ satisfies
$$\eqalign{{\ts{ 1\over4}}\sum_{k,j=0}^\infty\ &
{1\over\mua^{2k+2j+6}}\
\pak\paj
 \bar Y
+\sum_{k=0}^\infty {1\over\mua^{2k+4}}\pak \bar Y \cr
&+\half\sum_{k=0}^\infty \pak \bar Y\left(
{(2k+1)t_k\over\mua^4} + \cdots {t_0\over\mua^{2k+4}} \right)
+{1\over16\mua^4} \bar Y  =0 \ , }\eqn\eq
$$
which  yields the equations
$$\eqalign{
n=0~~ :&\qquad \sum_{k=0}^\infty (k+\half) t_k\pak \bar Y +{\ts
{1\over 16}}\bar Y =-\pa_0\bar Y
\ , \cr
n\geq1~~ :&\qquad \sum_{k=0}^\infty (k+\half) t_k\pa_{k+n}\bar Y +
{\ts{ 1\over4}}\sum_{k=1}^{n} \pa_{k-1}\pa_{n-k} \bar Y = - \pa_n \bar Y
\ . }
\eqn\eq
$$
We recognize these as Virasoro
constraints, namely
$$L_n \bar Y =- \pa_n \bar Y, \qquad n\geq 0 \ . \eqn\lybar
$$
These equations can be put in the standard form $L_n \bar Y =0$ by a
shift of $t_0$, which
makes explicit that~\lybar\ corresponds to an expansion about the point
$t_0=2$, $t_i=0$ for $i\geq1$. (Contrast this with the usual situation
in hermitian matrix models,
where $t_0\equiv x$ is a free parameter, and one other $t_m$,
 for some
 $m>0$, is
fixed to a  non-zero value.)
 This is an exceedingly trivial point, the $m=0$ unitary matrix
model,  which
corresponds naively  to a matrix model
with potential $V(U)=$ constant (the $m=1$ model has
$V=U+U^\dagger$);
 nevertheless it is  well defined,
since the integration is over a compact group.

The Virasoro constraints are a convenient starting point for extracting
useful information about the model. For instance, we can follow the
example of ref.~\dvv\ and derive from them recursion relations for the
correlation functions of scaling operators. For the $m=0$ model, these
take the form
$$\eqalign{&\langle\tau_n \prod_{k\in S} \tau_k \rangle_g = - \sum_{j\in
S}(j+\half) \langle\tau_{j+n}\prod_{k\neq j} \tau_k \rangle_g \cr
&-{\ts{1\over 4}}\sum_{j=1}^n \left\{ \langle\tau_{j-1}\tau_{n-j}
\prod_{k\in S} \tau_k\rangle_{g-1} +\half\!\! \sum _{S=X\cup Y \atop
g=g_1+g_2}
\langle\tau_{j-1} \prod_{k\in X}\tau_k\rangle_{g_1} \langle\tau_{n-j}
\prod_{k\in Y}\tau_n\rangle_{g_2} \right\} \ .  }\ee
$$
(The genus dependence of these correlation functions has been inserted
by hand, so as  to correspond to the conventional $1/N$ expansion.)
Note that the recursion relations are precisely sufficient to determine
all the correlation functions as pure numbers -- \ie, all the operators
are redundant. In a loose sense, therefore, this model is also
``topological.''

 There is also a  selection rule,
$$\sum_k k\, n_k =g-1  \qquad \Longleftrightarrow \qquad \sum_k k \, t_k \pak
 \left[ \log \bar  Y \right]_g =
(g-1) \left[ \log \bar Y\right]_g \ , \eqn\uselr
$$
which, like~\hselr, was  found  in ref.~\us. As a check,
 note that~\uselr\ and~\hselr\
are merely different linear combinations of
$L_0$ and the dilaton equation
$\sum_k n_k=2-2g$.\foot{We thank R. Dijkgraaf for pointing
this out to us.}
(In general,   one finds  the
selection rule for the $m^{th}$ multicritical model
 by eliminating $t_m$ between  the two equations.)

An important feature of~\lybar\ is  that because there is no $L_{-1}$ equation,
 the constant in $L_0$ is not determined simply by the requirement of closure
of the
Virasoro algebra. In ref.~\holl\ the constant is
 considered a free parameter of the
solution,
but our derivation singles out the value
${\ts{1\over16}}$.

There is another      difference between our results and
theirs, which we do not really understand. The constraints found here
 act directly on $\bar Y$, whereas
in ref.~\holl\  the partition function is a product of
{\it two} tau functions, each of which is annihilated separately by the
$L_n$'s.

\chapter{Symmetric matrices and unoriented surfaces.}

In this final section we discuss an interesting example of a model
which can  be solved only partially  by the methods we have been using:
  the real-symmetric matrix model.
This
\ref\bna{E.~Br\'ezin and H. Neuberger, \PRL {\bf 65} (1990) 2098.}
\ref\bnb{E.~Br\'ezin and H. Neuberger, \NP {\bf B350} (1991) 513.}
\ref\harris{G. Harris and E. Martinec, \PL {\bf 245B} (1990) 384.}
\ref\hneu{H. Neuberger, \PL {\bf 257B} (1991) 45.}
 has been investigated in several papers~[\bna,
\bnb, \harris, \hneu], and has a well-known geometric interpretation as the
sum over unoriented surfaces.
It appears that this model in an external field
 has not been  looked at before, so
 we permit ourselves to stray
somewhat from the subject  of Virasoro constraints and  discuss this
case in a fair amount of detail.
 Much of the analysis runs  parallel to that of the
hermitian matrices, and wherever possible we use the same notation

The starting point is the integral
$$\tilde\cz =\int\prod_{1\leq i\leq j\leq N} dM_{ij} \
 \exp N   \tr\!\!\left(XM-\gg M^3\right) \ ,
\eqn\symint
$$
for some real symmetric matrix $X$.
In this section we need to count
  powers of $N$ in the
topological expansion, so we include an $N$ in the exponent.

The derivation of the  Schwinger-Dyson equations from eqn.~\symint\
is more subtle than
for hermitian matrices  because the matrix elements of $M$ (or $X$) are
no longer independent, and when we differentiate
with respect to off-diagonal elements there is   an extra factor of 2. For
example,
$$ {\pa\over\pa M_{ij}} \tr \! (XM) =\left\{\eqaligntwo{  X_{ij}+X_{ji}&
 = 2X_{ij} &\qquad  i\neq j \cr
 \qquad X_{ii}&    &\qquad i=j }\right. \ . \ee
$$
As a consequence, the equations of motion have a complicated form when
written
 in terms of derivatives with
respect to the matrix elements of $X$. For the same reason we have to
be careful
when we change variables to eigenvalues, which requires evaluating such
quantities as $\pa^2 x_a /\pa X_{pq} \pa X_{rs}$. Remarkably, the
 final answer is both simple and familiar:
$$
 {\pa^2 \tilde \cz\over\pa x_a^2} + {1\over2}\sum_{b\neq a} {1\over
x_a-x_b}\left({\pa \tilde \cz\over \pa x_a}-{\pa \tilde \cz\over \pa x_b}
\right)  ={N^2\over g} x_a \tilde \cz \ . \eqn\ztilde
$$
However, there is a major difference between this and eqns.~\zhat, \zbar:
namely, the factor of $\half$ multiplying
 the sum. This
 small change makes an important  qualitative difference to
the  genus
expansion.

Before we solve~\ztilde,  it will prove
 convenient to rescale variables to $\la=4gx_a$.   Then,
 following the procedure of ref.~\us, we define
$\tilde\z=e^{N\tilde F}$ and rewrite~\ztilde\ as an equation for $\tilde\fa=\pa
\tilde F/\pa \la$,\foot{Note that $\tilde F$ is of order $N$, and
$\tilde F_a$ is of order 1.}
$$ \N{\pa\tilde\fa\over\pa \la} + \tilde
\fa^2 +{1\over2N}\sum_{b\neq a} {\tilde\fa-\tilde\fb\over \la-\lb}
={1\over 64g^4} \la \ . \eqn\ftilde
$$
In the spherical limit the first term is smaller by a factor of $1/N$,
and we can drop it. The resulting equation has the solution
$$\tilde\fa^{(0)} = {\nu_a \over 8g^2}+{1\over4 \nu_a} \left(\sig_1 -
\hat Z_a
 \right) \ , \eqn\fazero
$$
where it is useful to introduce the notation
$$\nu_a=\sqrt{\la+y}\ ,\qquad \sig_k =\N\sum_b
{1\over\nu_b^{ k}} \ , \qquad \hat Z_a =\N\sum_b {1\over \nu_a+\nu_b} \ ,
\ee
$$
 and $y$ is found from the equation
$y=-4g^2\sig_1$;
the superscript on $\tilde F_a$ refers to the order in $1/N$.
Integrating eqn.~\fazero, we find that the free energy is
$$ \N \tilde F^{(0)} = {1\over12g^2} \sig_{-3} +{1\over2}
\sig_1\sig_{-1}  +{g^2\over 6}\sigma_1^{~3} - {1\over 4N^2}
\sum_{b,c} \ln\left(\nu_b+\nu_c\right) \ . \ee
$$
We can attempt to solve~\ftilde\ order by order in $1/N$ to find the
corrections to
the free energy, as in ref.~\us.  Just write $\tilde F =\tilde
F^{(0)} + \N \tilde F^{(1)} + \cdots $,
insert into~\ftilde, and linearize. In the solutions of unitary and
hermitian matrices~[\us], the $1/N$ terms canceled and the leading
correction was O($1/N^2$). In the present case, this cancellation
does not occur, because of the factor
of $\half$; therefore the free energy
has a contribution of order $1/N$, as  one would  expect from topological
\ref\tHooft{G.~'t Hooft, \NP {\bf B72} (1974) 461.}
arguments~[\tHooft].

The resulting linear equation for $\tilde \fa^{(1)}$ is
$$\eqalign{\left(  {  \nu_a\over4g^2} +{1\over2 \nu_a} \left(\sig_1 -\hat Z_a
\right) \right) &\tilde \fa^{(1)}
+{1\over 2N}\sum_{b\neq a} {\tilde \fa^{(1)}-\tilde \fb^{(1)}\over
\la-\lb} =\cr
 -{1\over 32g^2 \nu_a}& +{1\over 16\nu_a^{3}} \left(\sig_1 -\hat
Z_a
 \right) -{1\over16N} \sum_b {1\over\nu_a^{2} (\nu_a+\nu_b)^2}
\ . } \eqn\ftlin
$$
The right-hand side here is  more complicated than
anything that arose in the unitary or hermitian cases. We have not
been able to solve this equation, and unfortunately  it seems probable that
 the  solution to~\ftlin\ cannot be written in closed form.
If this is true, then it is impossible even to write down the equations for
higher
$1/N$ corrections, let alone solve them.

On the other hand there is no  obstacle to
solving~\ftlin\  perturbatively in $g$ (actually, as  a double
power series expansion in $g$ and $1/N$). Rather than using diagrammatic
perturbation theory, we can work directly from the differential
equation~\ftilde.
But first,
note that we are principally interested in the Kontsevich-type integral
$$\cz =\int\cd M \exp N \tr\!\!\left(-\half AM^2-\gg M^3\right) = e^{NF}
\ ,
\eqn\symkont
$$
where $F$ and $\tilde F$ are related by $F=\tilde F -\tr A^3/12g^2$.
 Then one can show that $F_a=\pa F/\pa \la$ satisfies
$$ \N\left( {\pa\fa\over\pa\la}+{1\over 32g^2 \mu_a}\right) +\fa^2
+{\mu_a\over4g^2} \fa +{1\over2N}\sum_{b\neq a}{\fa-\fb\over \la-\lb}
+{Z_a\over 16g^2} =0 \ , \ee
$$
where $\mua$, $\la$, and $Z_a$ are defined  in section~2.
To derive the perturbation expansion, it is convenient to rearrange this into
the form
$$\fa=-{Z_a\over 4\mua} -{1\over 8N\mua^2}
-{4g^2\over\mua}\left({1\over2N}\sum_{b\neq a}{\fa-\fb\over
\la-\lb}+\fa^2+\N{\pa\fa\over\pa\la}\right) \ . \ee
$$

With some help from {\it Mathematica}, we find that the first few terms
of the solution at order $1/N$ are
$$ \eqalign{\N F^{(1)} =& -{1\over4N} \sum_b\ln\mu_b +g^2\left( \half s_1 s_2
-{1\over4N^2}\sum_{b,c} {1\over \mu_b\mu_c(\mu_b+\mu_c)}\right) +\cr
& g^4\left(s_1^{~2} s_4
+\half s_1 \N\sum_{b,c}{1\over\mu_b^2\mu_c^2(\mu_b+\mu_c)}
\ -\right. \cr
&\qquad \left. {1\over 6N^3}\sum_{a,b,c}{1\over \mu_a\mu_b\mu_c (\mu_a+\mu_b)
(\mu_b+\mu_c)(\mu_c+\mu_a)} \right) + {\rm O}(g^6) \ , } \eqn\freee
$$
where we have used the notation  $s_k=\N\sum_b{1\over \mu_b^{k}}$.
This can be written in a more concise form in terms of the
``shifted'' eigenvalues $\nu_a \ (\equiv\sqrt{\mu_a^2 +y})$,
$$\eqalign{\N F^{(1)} = -{1\over4N}& \sum_b\ln\nu_b -{g^2\over 4N^2}\sum_{b,c}
{1\over \nu_b\nu_c(\nu_b+\nu_c)}\ -\cr
 &{g^4\over6N^3}\sum_{a,b,c}{1\over
\nu_a\nu_b\nu_c (\nu_a+\nu_b)
(\nu_b+\nu_c)(\nu_c+\nu_a)} +{\rm O}(g^6)
\ .} \eqn\freea
$$
To recover~\freee\ from this, use the perturbative expansions of
$\nu_a$,
$$ \nu_a = \mua\left( 1+\half y{1\over \mua^{2}} + \cdots \right) \ ,
\ee
$$
and of $y$,
$$ y=-4g^2 \N\sum_b{1\over(\mu_a^{2}+y)^{1/2}}= -4g^2 \left( s_1 -\half y
s_3 +\cdots \right) \ . \eqn\freeb
$$
This latter equation can be solved iteratively to give $y$ in terms of
the $s_k$'s.

The  notable feature of~\freea\ is that it  cannot be expressed  in
terms of just the traces $s_{2k+1}$, in contrast to the unitary and
hermitian cases.  In fact, it would  appear that at
each order  in $g^2$ we encounter new  and more complicated invariants of
the eigenvalues. Because the solution in its concise
form~\freea\ is so simple, one is tempted to  guess the general term at
order $g^{2n}$. But this simplicity is misleading, and even at O$(g^6)$
we do not know the form of the solution.
 The obvious ansatz,
 namely
$${1\over N^4}\sum_{a,b,c,d} {1\over \nu_a\nu_b\nu_c\nu_d
(\nu_a+\nu_b) \cdots (\nu_c+\nu_d)} \ , \ee
$$
can be ruled out  on dimensional grounds, as we need a term of
order $1/\nu^9$, not $1/\nu^{10}$.

We finish by asking whether it is possible to
 find Virasoro-type constraints for
unoriented surfaces. We have the first ingredient, namely
the Schwinger-Dyson equations~\ftilde. Next we must identify how the
integral~\symkont\ depends on the eigenvalues of $A$, and ``change
variables.''
In the previous examples we were lucky: a perturbative analysis
showed that the integrals were functions solely of the $t_k$'s, and the
change of variables was simple. In the present
case,  we
don't even know the complete set of variables; but we do know that it is
complicated, and the change of
 variables would likely  present a thorny challenge.
For these reasons, we think that further progress in this direction is
impossible.

\chapter{Discussion.}

In this paper we have demonstrated an efficient method of deriving
Virasoro constraints satisfied by the Kontsevich model, which we
generalized
 to the unitary matrix model~\umat.
Now we ask whether this has revealed anything about the geometrical
interpretation of unitary matrices.
To answer this, we consider first the hermitian case~\kint, which is
well understood.

There is an important point about the Kontsevich model which we
wish to emphasize. The great majority of papers on matrix models involve
studying them near their critical points, which is where the perturbation
expansion diverges and the Feynman diagrams have an interpretation as
continuum surfaces discretized by large numbers of infinitesimal
triangles.
This is not the situation in the Kontsevich model, where one interprets
a Feynman diagram of low order in perturbation theory as a surface with
a finite number of punctures (equal to the number of faces on the
diagram). This is a topological picture, in which  all surfaces of given
genus and number of punctures are represented by a single
``ribbon graph''; and it yields topological information, \ie, the
intersection numbers. In addition,   the integral~\kint\ also
possesses a conventional  critical point, studied in ref.~\us, with an
 interpretation
in terms of continuum
surfaces.

This situation is reminiscent of the well-known Penner model, whose
perturbative ``topological'' limit encodes  the virtual Euler
characteristic of surfaces~[\penner,~\harer,~\bowditch],
\ref\distler{J. Distler and C. Vafa, \MPL {\bf A6} (1991) 259.}
whereas the continuum limit corresponds to a  $c=1$ string in a
compactified target space of a particular radius~[\distler].

Now we conjecture  that the unitary matrix model of
section~3 is a third example of this type.
The integral certainly possesses the two appropriate limits. From the
``perturbative'' limit we derive Virasoro constraints. The work of
Hollowood {\it et al.}~[\holl] gives an independent confirmation that
these are the correct
 constraints to  describe the multicritical unitary matrix models, which
we wish to interpret as some ensemble of surfaces coupled to matter. The
non-trivial step is to suggest  that this structure arises here because the
integral~\umat\ has a representation, analogous to that of the
Kontsevich integral, in terms of the ``moduli space''
of the same (unknown) ensemble. However, we leave open the question of
what
this ensemble might be.

It is very disappointing that we were unable to repeat the analysis for
symmetric matrices, but  the moduli space of unoriented
surfaces is  known to be
\ref\cpb{C. P. Burgess and T. R. Morris, \NP {\bf B291} (1987) 285.}
a lot more complicated than that of oriented surfaces.\foot{For example,
see ref.~\cpb.}
Nevertheless, it has been  instructive to study this model
for  the subtleties  it reveals in
  the hermitian and unitary matrix models.

As a  possible generalization of this work, one could try to extend these
results to the $d=-2$ external field model~[\Kaz]. This would
not be trivial, however, as that model was solved by a different method.

The original motivation for the research that led to this paper was to
find a geometrical picture
 for unitary matrix models.
  This is a goal that has
\ref\CDMA{\v C. Crnkovi\'c and G. Moore, \PL {\bf 257B} (1991) 322.}
 long been  sought by many others (\eg~[\CDMB,~\Neu,~\CDMA]).
Unfortunately
 a satisfactory answer eludes us still,
 but we offer our insights in
the hope that others can continue this work to completion.

\ref\mmm{A.~Marshakov, A.~Mironov, and A.~Morozov, ``On equivalence of
topological and quantum 2d gravity,''  preprint HU-TFT-91-44,
FIAN/TD/04-91, ITEP-M-4/91.}
\ref\newkont{M.~Kontsevich, ``Intersection theory on the moduli space of
curves and the matrix Airy function,''
Max Planck Institute preprint MPI/91-77.}
\noindent {\it Note added}: While this paper was being prepared, we
learned
that A.~Marshakov, A.~Mironov, and A.~Morozov~[\mmm], and
M.~Kontsevich~[\newkont],
have independently derived these same results for hermitian matrices.
\ref\kmmmz{ S.~Kharchev, A.~Marshakov, A.~Mironov, A.~Morozov, and
A.~Zabrodin, ``Unification of all string models with $c<1$,'' preprint
FIAN/TD/09-91, ITEP-M-8/91; ``Towards unified theory of 2$d$ gravity,''
preprint FIAN/TD-10/91, ITEP-M-9/91.}
In addition, S.~Kharchev, A.~Marshakov, A.~Mironov, A.~Morozov, and
A.~Zabrodin~[\kmmmz] have recently proposed a generalized Kontsevich model to
describe all multi-matrix models.

\ack
We are  grateful to the authors of ref.~\holl\ for sharing their
 results with us prior to publication, and in particular to Tim Hollowood and
Andrea Pasquinucci for valuable discussions. We would also
like to thank Ulf Danielsson, Robbert Dijkgraaf, Jacques Distler, Miguel
Martin-Delgado, Herman Verlinde,
  and Ed Witten
 for frequent help and explanations  along the
way. M.~N. is especially indebted to Mark Doyle for many useful
conversations, and in particular for his thorough
explanation of the paper of Kontsevich.

\refout
\bye